\documentstyle[preprint,aps]{revtex}
\begin{document}
\draft
\preprint{DOE/ER/40427-22-N93}
\title{
Contribution of Quark-Mass-Dependent Operators to \\
Higher Twist Effects in DIS}
\author{
Su Houng Lee\footnote{On leave of absence from
Department of Physics, Yonsei University, Seoul 120-749, Korea}
}
\address{
Department of Physics, FM-15 \\
University of Washington \\
Seattle, WA 98195}
\date{\today}
\maketitle
\begin{abstract}
We look at the contribution of Quark-Mass-dependent twist-4 operators
to the second moments of spin averaged structure functions and
the Bjorken sum rule.  Its contribution is non-negligible in the former case
due to large Wilson coefficients.  We also discuss the values of the twist-
4 spin-2 nucleon matrix element within present experimental constraints.

\end{abstract}
\pacs{PACS number: 13.60.Hb}
\newpage

\section{Introduction}
\setcounter{equation}{0}

Recent precision spin averaged lepton-hadron Deep Inelastic Scattering(DIS)
data at CERN \cite{NMC92,NMC1,NMC2,BCDMS} and SLAC\cite{SLAC}
provide us with  useful information about
higher twist effects in spin averaged structure functions($F_2$ and $F_L$).
 From these data,  we have previously\cite{CHKL} extracted the higher
twist effect in the second moments of the structure functions and  obtained
a model independent constraints
on the nucleon matrix elements of the twist-4 spin-2 operators.
These matrix elements provide useful information about correlations between
partons inside the nucleon.

In addition, polarized DIS provides crucial information on the spin structure
of the nucleon\cite{EMC,E142,SMC}.
The lowest moment of this spin structure function is
related to  the nucleon matrix element of the axial current.  To actually
estimate the quark contribution of the proton spin and to
understand
differences in experimental measurements at different scales,
it is necessary to estimate the higher twist-effects\cite{EK93,CR93,BBK90}.

Higher twist operators appearing in both spin averaged moments
and the Bjorken sum rule have been considered for some time
\cite{P80,SV82,JS,EFP82}.
In this work, we study the contribution of twist-4 operators, that are
proportional to the quark masses.   These operators have been neglected so
far because of negligible up and down quark masses compared to the hadronic
scale.
 For the Bjorken sum
rule, the additional twist-4 operator is $\langle m^2 \gamma_5 \gamma_\alpha q
\rangle$ and indeed negligible.
However, for the second moment of spin averaged structure function, the
quark-mass dependent operator $\langle \bar{q} D_\alpha D_\beta m q \rangle$
contributes
with a large Wilson coefficient
compared to other twist-4 operators and can not be neglected.

In section 2, we will first discuss  the Bjorken sum rule.
The relevant Operator Product Expansion (OPE)
for the time ordered vector current is derived
using the method based on
the Fock-Schwinger gauge, which has been  used  extensively in QCD sum rule
calculations\cite{SVZ} and DIS\cite{SV82,BK87}.
In section 3, we will discuss second moments for the spin averaged
structure function and discuss the importance of quark mass dependent
operators in relation to the recent estimates for the value of total
twist-4 effects.   We will also discuss the values of twist-4 spin-2 matrix
elements within present experimental constraints.

\section{Bjorken sum rule}

\noindent{\bf OPE}

\vspace*{0.3cm}

Let us consider the  time ordered correlation of the electromagnetic
current.

\begin{eqnarray}
T_{\mu \nu}(q)=i \int d^4x e^{iqx} \langle T[ j^{em}_\mu (x) j^{em}_\nu]
\rangle
\end{eqnarray}

The perturbative part is symmetric in $\mu \nu$ and the lowest antisymmetric
part contributing to the Bjorken sum rule is related to the axial current.
A simple
way to obtain the twist-4 spin-1 contribution, is to make use of the
Fock-Schwinger gauge for the external gauge fields.  We will follow similar
steps given in ref.\cite{SV82}.

\begin{eqnarray}
\label{first}
T_{\mu \nu}^A(q)=- \int d^4x e^{iqx} [
\langle \bar{q} \gamma_\mu S^{(2)}(x,0) \gamma_\nu Q^2 q \rangle +
x^\alpha \langle
\bar{q} \stackrel{\leftarrow}{D}_\alpha
\gamma_\mu S^{(1)}(x,0) \gamma_\nu Q^2 q \rangle
\nonumber \\[12pt]
+ \frac{1}{2} x^\alpha  x^\beta
\langle \bar{q} \stackrel{\leftarrow}{D}_\alpha \stackrel{\leftarrow}{D}_\beta
\gamma_\mu S^{0}(x,0) \gamma_\nu Q^2 q \rangle]
-[ \mu \leftrightarrow \nu ].
\end{eqnarray}
Here, the $S^{(i)}(x,0)$ are quark propagators in external gauge field with
$i$  being the sum of the dimensions of external gauge field and quark mass.
These are summarized in Appendix A.
There are only two independent twist-4 operators contributing to this
order.

\begin{eqnarray}
U_\alpha=\bar{q} g \stackrel{\sim}{G}_{\alpha \beta} \gamma_\beta Q^2 q ,
{}~~~~ L_\alpha=\bar{q} \gamma_\alpha \gamma_5 m^2  Q^2 q ,
\end{eqnarray}
where, $\stackrel{\sim}{G}_{\alpha \beta}=\frac{1}{2}
\epsilon_{\alpha \beta \mu \nu}
G_{\mu \nu}$ and conventions for $\epsilon_{0123}$ and $\gamma_5$ follow that
of Muta\cite{Muta}.  $Q$ is the flavor SU(2) charge matrix.
To express the second and third
term in eq.({\ref{first}) in terms of these independent operators,
we have to extract the spin-1 part of the
following operators.
\begin{eqnarray}
\bar{q} \stackrel{\leftarrow}{D}_\alpha [\gamma_\beta, \gamma_\delta] m q, ~~~
\bar{q} \stackrel{\leftarrow}{D}_\alpha
\stackrel{\leftarrow}{D}_\beta \gamma_5 \gamma_\delta q
\end{eqnarray}
This can be done\cite{SV82}
by assuming the spin-1 part of the  operators to be,
\begin{eqnarray}
{\cal O}_{\alpha \beta \delta}-{\cal O}_{\alpha \beta \delta}^{spin-3}=
g_{\alpha \beta} X_\delta+g_{\beta \delta} Y_\alpha + g_{\alpha \delta}
Z_\beta+i\epsilon_{\alpha \beta \delta  \sigma} U_\sigma.
\end{eqnarray}
Taking all traces and using the following identities,
\begin{eqnarray}
D_\alpha \gamma_\alpha q & =& -imq \nonumber \\[12pt]
\epsilon_{\alpha \beta \mu \nu } D_\mu \gamma_\nu q
& =&-i\gamma_\alpha \gamma_5 D_\beta q +i\gamma_\beta \gamma_5 D_\alpha q
-m\gamma_5(\gamma_\alpha \gamma_\beta-g_{\alpha \beta}) q
\end{eqnarray}
We find,
\begin{eqnarray}
\bar{q} \stackrel{\leftarrow}{D}_\alpha [\gamma_\beta ,\gamma_\delta] m q
& = & \frac{2}{3} \epsilon_{\alpha \beta \delta  \sigma}  L_\sigma
\nonumber \\[12pt]
\bar{q} \stackrel{\leftarrow}{D}_\alpha \stackrel{\leftarrow}{D}_\beta
\gamma_5 \gamma_\delta q
& =& g_{\alpha \beta}\frac{5}{18}( L-U)_\delta +
    g_{\alpha \delta} \frac{1}{18}(U- L)_\beta +
    g_{\beta \delta} \frac{1}{18}(U- L)_\alpha
\end{eqnarray}
Using these results, the final answer including the leading order result
gives,
\begin{eqnarray}
T_{\mu \nu}^A(q)=-\frac{i}{q^2} \epsilon_{\mu \nu \alpha \beta}q_\alpha
2V_\beta
-\frac{i}{q^4} \epsilon_{\mu \nu \alpha \beta} q_\alpha
( \frac{16}{9} U_\beta +\frac{20}{9} L_\beta)
\end{eqnarray}
where the leading operator is,
\begin{eqnarray}
V_\beta=\bar{q} \gamma_\beta \gamma_5 Q^2 q.
\end{eqnarray}

\vspace*{0.5cm}

\noindent{\bf Correction to the Bjorken sum rule}

\vspace*{0.3cm}

The contribution of $L_\alpha$ is easy to estimate.  One should note that
to a good approximation, $L_\alpha=m^2 V_\alpha$, where $m^2$ is the
average quark mass squared.   So the contribution of this operator compared
to the leading result would just be $\frac{10}{9}\frac{m^2}{Q^2}$.
Since $m$ is
$5 \sim 10 $ MeV$^2$, its contribution is indeed negligible  down to very
small $Q^2$.

\section{Second moments}

\noindent{\bf OPE}

\vspace*{0.3cm}

To obtain the second moment, we have to consider the OPE to dimension 6 such
that  the contribution, where one quark is connected, is given by,
\begin{eqnarray}
\label{second}
T_{\mu \nu}^c(q)& = & - \int d^4x e^{iqx} [
\langle \bar{q} \gamma_\mu S^{(3)}(x,0) \gamma_\nu Q^2 q \rangle +
x^\alpha \langle \bar{q} \stackrel{\leftarrow}{D}_\alpha
\gamma_\mu S^{(2)}(x,0) \gamma_\nu Q^2 q \rangle
\nonumber \\[12pt]
& & + \frac{1}{2} x^\alpha  x^\beta
\langle \bar{q} \stackrel{\leftarrow}{D}_\alpha \stackrel{\leftarrow}{D}_\beta
\gamma_\mu S^{1}(x,0) \gamma_\nu Q^2 q \rangle]
+ \frac{1}{3} x^\alpha  x^\beta x^\delta
\langle \bar{q} \stackrel{\leftarrow}{D}_\alpha
\stackrel{\leftarrow}{D}_\beta \stackrel{\leftarrow}{D}_\delta
\gamma_\mu S^{0}(x,0) \gamma_\nu Q^2 q \rangle]
\nonumber \\[12pt]
& & +[ \mu \leftrightarrow \nu ]
\end{eqnarray}

There are four independent twist-4 operators appearing to this
order.

\begin{eqnarray}
\label{op}
A_{\alpha \beta} & = & g \bar{q} [ D_\mu, G_{\alpha \mu}]\gamma_\beta Q^2 q
{}~~~=~~g^2 (\bar{q} \gamma_\alpha t^a q)( \bar{q} \gamma_\beta t^a Q^2 q)
\nonumber \\[12pt]
B_{\alpha \beta} & = & g \bar{q} \{ i D_\alpha,
\stackrel{\sim}{G}_{\beta \mu } \} \gamma_5
\gamma_\mu Q^2 q
\nonumber \\[12pt]
C_{\alpha \beta} & = & \bar{q}  D_\alpha D_\beta m Q^2 q
\nonumber \\[12pt]
D_{\alpha \beta} & = & g \bar{q} [ D_\alpha, G_{\mu \beta}]\gamma_\mu Q^2 q
\end{eqnarray}
Here, the operators are assumed to be symmetric and traceless with respect
to the lorentz indices.   $t^a$ are
the generators of SU(3) color matrix normalized to $Tr(t^a)^2=\frac{1}{2}$.

The last operator does not contribute to the final answer.   Similar steps
have to be taken as before to reduce eq.(\ref{second}) into the above
independent form.  Useful identities needed in the intermediate stages
are given in Appendix B.
The final answer is,

\begin{eqnarray}
T_{\mu \nu}^c &  =& \frac{1}{x^2 Q^2} [d_{\mu \nu}(\frac{5}{8}A+\frac{1}{8}B
         -\frac{13}{4}C)   \nonumber \\[12pt]
 & & + e_{\mu \nu}(\frac{1}{4}A-\frac{3}{4}B-\frac{18}{4}C)],
\end{eqnarray}
Here, $A,B,C$ are the spin-independent part of the spin-averaged nucleon
matrix
elements of the twist-4 spin-2 operators in eq.(\ref{op}) such that
$ \langle A^i_{\alpha \beta} \rangle_N=(p_\alpha
p_\beta-\frac{1}{4} M_N^2 g_{\alpha
\beta} ) A^i$ with $A^1=A,A^2=B,A^3=C$.
The polarization tensors are defined as $e_{\mu \nu}=g_{\mu \nu}-q_\mu q_\nu
/q^2$ and $d_{\mu \nu}=-p_\mu p_\nu q^2/(p \cdot q)^2 +(p_\mu q_\nu +p_\nu
q_\mu)/p \cdot q-g_{\mu \nu}$ with $Q^2=-q^2$($p_\mu$ is a 4-momentum of the
nucleon with $p^2=M_N^2$).
There is another contribution to twist-4 part coming from a disconnected
four quark operator.   This part does not have
any additional quark mass dependent term,
such that the full twist-4 part in lowest order in $\alpha_s$ is,
\begin{eqnarray}
\label{final}
T_{\mu \nu}^{twist-4} =\frac{1}{x^2 Q^2}d_{\mu \nu}  M_4+T_{\mu \nu}^c ,
\end{eqnarray}
where $M_4$ is defined from
\begin{eqnarray}
M_4(p_\alpha p_\beta-\frac{1}{4} M_N^2 g_{\alpha \beta}) & = &
 \langle (\bar{q} g^2 \gamma_\alpha \gamma_5 \tau^a q)
(\bar{q} g^2 \gamma_\beta \gamma_5 \tau^a q) \rangle_N
\end{eqnarray}
Eq.(\ref{final}) was obtained without the quark mass dependent
operators in ref.(\cite{SV82,JS})

\vspace*{0.5cm}

\noindent{\bf Estimate of C }

\vspace*{0.3cm}

C is related to the second moment of the structure function
$e(x)$\cite{JJ}.  At present, there is no direct experimental way of
measuring $e(x)$.   To estimate C,
we will make the following approximation.

\begin{eqnarray}
\label{ratio}
 \langle \bar{q} D_\alpha D_\beta Q^2 m q \rangle_N
\sim
- \frac{4}{9}P^u_\alpha P^u_\alpha \langle m_u \bar{u} u \rangle_N
-\frac{1}{9}P^d_\alpha P^d_\alpha \langle m_d \bar{d} d \rangle_N
\end{eqnarray}
Here, $P^u(P^d)$ is the average momentum carried by each $u(d)$ quark inside
the nucleon, which should be $\frac{1}{6}$ of the nucleon momentum $P$ such
that the total momentum carried by three quarks is roughly $\frac{1}{2}P$.
Assuming $\langle m_u \bar{u} u \rangle \sim 2 \langle
m_d \bar{d} d \rangle$ in the proton
and using
$\Sigma_{\pi N} \sim 45 \pm 10$ MeV\cite{GLS}, we have in the covariant
normalization
\begin{eqnarray}
\langle m_u \bar{u} u \rangle_N=2\langle
m_u \bar{d} d \rangle_N =0.06 {\rm~~~GeV}^2.
\end{eqnarray}
This implies
\begin{eqnarray}
\label{C}
\langle \bar{q} D_\alpha D_\beta Q^2 m q \rangle_N \sim
(P_\alpha P_\beta -\frac{1}{4} M_N^2 g_{\alpha \beta})
\times (-0.0008 ) {\rm ~~GeV}^2. {\rm ~~~~proton}
\end{eqnarray}
So $C=-0.0008 $ GeV$^2$ for the proton.

\vspace*{0.5cm}

\noindent{\bf Experimental Data on twist-4 effects in $F_2$}

\vspace*{0.3cm}

We have recently extracted the twist-4 effects in the second moment of
the structure function $F_2$\cite{CHKL}.   Assuming,
\begin{eqnarray}
F_{2,L}(x,Q^2) = F_{2,L}^{\tau=2} (x,Q^2) + {1 \over Q^2} F_{2,L}^{\tau =4}
(x,Q^2),
\end{eqnarray}
where $F_{2,L}^{\tau=2}$  takes into account the target mass correction.
We found that at $Q^2=5$ GeV$^2$

\begin{eqnarray}
\label{F2}
\int_0^1 F_2^{\tau =4}(x) dx
& = &  0.005 \pm  0.004     {\rm GeV}^2 {\rm ~~~~~(proton)} \nonumber \\
& = & \frac{1}{2} (M_4 +\frac{5}{8}
A+\frac{1}{8}B -\frac{13}{4}C  ),
\end{eqnarray}
where the second line shows its relation to the matrix elements including
$C$.
The quoted value was obtained from estimates of $F_2^{\tau =4}(x)$
in ref.\cite{VM}, which in turn was obtained by fitting
the BCDMS and SLAC data on the hydrogen target.   We also quoted a value for
the neutron which was based on the analysis of NMC group for higher twist
effect in $F_2^n/F_2^p=2F_2^d/F_2^p-1$\cite{NMC2}.  This ratio was obtained
 from the measurement in the deuteron and proton targets without taking into
account any nuclear effects in the deuteron structure function.
However, recent analysis showed\cite{EFGLS,BGNPZ} that
the nuclear effect is not
negligible and accounted for almost all the deviation from the
the perturbative ratio.   So we will not quote the neutron value and
hereafter discuss the case of the proton only.

Using the value for C From eq.(\ref{C})  and substituting in
eq.(\ref{F2}), we see that the
contribution from C accounts for  26 \% of the
experimentally estimated twist-4 effects and can not be neglected.

\vspace*{0.5cm}

\noindent{\bf Experimental Data on twist-4 effects in $F_L$}

\vspace*{0.3cm}

In ref.\cite{CHKL}, we made similar estimates for the twist-4 effect in the
second moment
of $F_L$.  This was based on the
SLAC data \cite{SLAC} analysed in ref.\cite{GMMPS91}.
At $Q^2=5$ GeV$^2$ we found
\begin{eqnarray}
\label{FL}
\int_0^1
F_L^{\tau=4}  dx
 & = &  \int_0^1 8 \kappa^2 F_2^{LT}(x) dx =
             0.035  \pm  0.012     {\rm GeV}^2 {\rm ~~~~(Proton)}
 \nonumber  \\
& = & \frac{1}{2} ( \frac{1}{4}
A-\frac{3}{8} B-\frac{18}{4} C ),
\end{eqnarray}
where the second line shows its relation to matrix elements.

Using the value for C From eq.(\ref{C}) and
substituting in eq.(\ref{FL}), we see that in this case,
contribution from the quark mass dependent operator accounts for  only
$5 \% $.

\vspace*{0.5cm}

\noindent{\bf An estimate of the proton matrix element of $ M_4, A$ and $ B$}

\vspace{0.3cm}

Let us now estimate the value of $M_4$, $A$ and $B$ for the proton.
We will assume $C=-0.0008$ GeV$^2$.  There are only two constraints
(eq.(\ref{F2}) and  eq.(\ref{FL})) and 3 parameters to determine.
However, there is one positivity constraint(\cite{SV82}) for the four quark
operators.  The proof is simple.  Suppose the twist-4 four quark
operator to be,
\begin{eqnarray}
{\cal O}_{\mu \nu}=(\bar{q} \Gamma_\mu q)(\bar{q} \Gamma_\nu q)-
\frac{1}{4} g_{\mu \nu} (\bar{q} \Gamma_\alpha q)(\bar{q} \Gamma_\alpha q),
\end{eqnarray}
then for $\langle \mbox{\boldmath $p$}=0 | {\cal O}_{00} |
\mbox{\boldmath $p$}=0 \rangle$,
the matrix element
can be written as the sum of full squares:
\begin{eqnarray}
\langle \mbox{\boldmath $p$}=0 |[\frac{3}{4}(\bar{q} \Gamma_0 q)^2+\frac{1}{4}
(\bar{q} \Gamma_i q)^2] | \mbox{\boldmath $p$}=0 \rangle
\end{eqnarray}
Hence  both $M_4$ and $A$ should be positive.  Since both come from four
quark operators, we expect them to be similar in magnitude.  We will
vary $M_4$ from $0$ to $2 \times A$ and solve for $A$ and $B$.
The result is summarized in Table 1.

Our analysis suggest that:

\begin{enumerate}

\item The magnitude of the quark gluon mixed operator $B$ is relatively
large compared to the four quark operators at the 5 GeV$^2$ scale.  The former
is in the range of -(350$\sim$450 MeV)$^2$ and the latter (100 $\sim$ 200
MeV)$^2$,
consistent with previous estimates\cite{CHKL}.

\item C is negligible compared to B but not so compared to A.  For the
transverse moment in eq.(\ref{F2}), the
Wilson coefficient of B is small and its
contribution is largely cancelled by that of A.  Therefore, the contribution
 from C, which comes with a large Wilson coefficient,
is a sizable part of the estimated twist-4 contribution and can not be
neglected.

\end{enumerate}

\section{Conclusion}

We have considered the contribution of quark-mass-dependent twist-4
operators to the
Bjorken sum rule and momentum sum rule.  In the Bjorken sum rule only
operators proportional to $m^2$ appear.  This is easily seen because the
operator linear in $m$ would be $\bar{q} m \gamma_5 D_\alpha q$ which
vanishes when using the equations of motion.  So the correction to the
leading part is
$\frac{m^2}{Q^2}$ and negligible.

However, for the momentum sum rules,
$\bar{q} D_\alpha D_\beta m q$ contributes with a large Wilson coefficient
and becomes a non-negligible correction to other twist-4 effects.

\vspace{1cm}

\centerline{\bf Acknowledgment}

We would like to Thank V. M. Braun for useful comments.
This work was supported in part by the U.S. Department
 of Energy under grant DE-FG06-88ER40427.

\newpage

\centerline{\bf Appendix A}
\setcounter{equation}{0}
\renewcommand{\theequation}{A.\arabic{equation}}
In this appendix, we summarize the form of quark propagator in external gauge
field including mass dependent terms relevant for the moment
calculations.   We will group them together according to the dimension of
external gauge field plus the dimension of quark mass.
\begin{eqnarray}
S(p)=S(p)^{(0)}+S(p)^{(1)}+S(p)^{(2)}+S(p)^{(3)}.
\end{eqnarray}
Such that,
\begin{eqnarray}
S(p)^{(0)} & = & \frac{1}{p^2} p_\alpha \gamma_\alpha,  \nonumber \\[12pt]
S(p)^{(1)} & = & \frac{1}{p^2} m,  \nonumber \\[12pt]
S(p)^{(2)} & = & -\frac{1}{p^4} p_\alpha
\stackrel{\sim}{G}_{\alpha \beta} \gamma_\beta
   \gamma_5 +\frac{1}{p^4} m^2 p_\alpha \gamma_\alpha,  \nonumber \\[12pt]
S(p)^{(3)} & = & \frac{2}{3} \frac{1}{p^6}g[p^2 D_\alpha G_{\alpha \beta}
\gamma_\beta- p_\sigma \gamma_\sigma D_\alpha G_{\alpha \beta} p_\beta-
p_\sigma D_\sigma p_\alpha G_{\alpha \beta} \gamma_\beta \nonumber \\[12pt]
& & -3i
p_\sigma D_\sigma p_\alpha \stackrel{\sim}{G}_{\alpha \beta}
\gamma_\beta \gamma_5]
- \frac{1}{2p^4} m g G_{\alpha \beta} \sigma_{\alpha \beta}
\end{eqnarray}

\vspace{20pt}
\centerline{\bf Appendix B}
\setcounter{equation}{0}
\renewcommand{\theequation}{B.\arabic{equation}}
In this appendix, we summarize useful identities that are needed to reduce
eq.(\ref{second}) into its finial form with independent matrix elements.
We will only look at the  spin-1 part of the following operators.
\begin{eqnarray}
\bar{q}  \stackrel{\leftarrow}{D}_\alpha
\stackrel{\leftarrow}{D}_\beta \stackrel{\leftarrow}{D}_\gamma
\gamma_\delta q & =
g_{\alpha \beta}\frac{i}{32}[2D-A-3B-2C]_{\gamma \delta}+
g_{\alpha \gamma}\frac{i}{32}[4D+5A-3B-2C]_{\beta \delta} \nonumber \\
 & + g_{\alpha \delta}\frac{i}{32}[-2D-A+B+6C]_{\beta \gamma}+
g_{\beta \gamma}  \frac{i}{32}[2D-A-3B-2C]_{\alpha \delta} \nonumber \\
 & +g_{\beta \delta}  \frac{i}{32}[4D+A+B+6C]_{\alpha \gamma}
+ g_{\gamma \delta} \frac{i}{32}[-2D-A+B+6C]_{\beta \alpha},
\end{eqnarray}
where we have neglected the terms proportional to $\epsilon$ tensor.

\begin{eqnarray}
\label{B2}
\bar{q}  \stackrel{\leftarrow}{D}_\alpha \stackrel{\leftarrow}{D}_\beta
\stackrel{\leftarrow}{D}_\gamma \gamma_\delta \gamma_5 q & =
 \epsilon_{\sigma \beta \gamma \delta} \frac{1}{16}
[-2B-A-D+4C]_{\sigma \alpha} +
 \epsilon_{\alpha \sigma  \gamma \delta} \frac{1}{16}
[-2A-2D+8C]_{\sigma \beta} \nonumber \\[12pt]
& + \epsilon_{\alpha \beta \sigma \delta} \frac{1}{16}
[-2B-A-D+4C]_{\gamma \sigma },
\end{eqnarray}
where we have neglected the terms proportional to $g$.  Using eq.(\ref{B2}),
we can show,
\begin{eqnarray}
E \equiv \bar{q} \{ i D_\sigma
\stackrel{\sim}{G}_{\sigma \alpha} \} \gamma_5 \gamma_\beta q=
[-B-A-D+4C]_{\alpha \beta}
\end{eqnarray}

Other straightforward expansion are,
\begin{eqnarray}
\bar{q} i \stackrel{\leftarrow}{D}_\alpha
\stackrel{\sim}{G}_{\beta \gamma} \gamma_5 \gamma_\delta q & =
g_{\alpha \gamma} \frac{1}{16}[3E+B]_{\beta \delta}
-g_{\alpha \beta} \frac{1}{16}[3E+B]_{\beta \delta} \nonumber \\[12pt]
& + g_{\beta \delta} \frac{1}{16}[3B+E]_{\beta \delta}
-g_{\gamma \delta} \frac{1}{16}[3B+E]_{\alpha \beta } ,
\end{eqnarray}
and,
\begin{eqnarray}
\bar{q} D_\alpha G_{\beta \gamma} \gamma_\delta q & =
g_{\alpha \beta} \frac{1}{8}[-3A-D]_{\gamma \delta}
-g_{\alpha \gamma} \frac{1}{8}[-3A-D]_{\beta \delta} \nonumber \\[12pt]
& + g_{\beta \delta} \frac{1}{8}[3D+A]_{\beta \delta}
-g_{\gamma \delta} \frac{1}{8}[3D+A]_{\alpha \beta } .
\end{eqnarray}

\newpage

\centerline{
\begin{tabular}{||c|c|c||}   \hline
{}~~~ $M_4$~~~ & ~~~ $A$ ~~~ & ~~~ $B$ ~~~
\\    \hline \hline
$0$  &  0.041 GeV$^2$  & -0.148 GeV$^2$
\\ \hline
$A=0.017$ GeV$^2$ &  0.017 GeV$^2$  & -0.165 GeV$^2$
\\ \hline
$2A=0.022$ GeV$^2$ &  0.011 GeV$^2$ & -0.169 GeV$^2$  \\ \hline
\end{tabular} }

\vspace{0.3cm}

\centerline{Table 1:Proton matrix elements of $M_4$, A and B.}

\end{document}